\documentclass[showpacs,superscriptaddress,preprintnumbers,amsmath,amssymb]{appolb}
\usepackage{epsfig}
\usepackage{graphicx}
\usepackage{dcolumn}
\usepackage{bm}
\usepackage{subfigure}
\usepackage{amsmath}

\begin{document}
\title{Fluctuations and correlations in Polyakov loop extended chiral fluid dynamics%
\thanks{Presented at Strangeness in Quark Matter 2011}%
}
\author{Christoph Herold and Marcus Bleicher
\address{Institut f\"ur Theoretische Physik, Goethe-Universit\"at, Max-von-Laue-Str.~1,
 60438 Frankfurt am Main, Germany}
\address{Frankfurt Institute for Advanced Studies (FIAS), Ruth-Moufang-Str.~1, 60438 Frankfurt am Main, Germany}
\and
 Marlene Nahrgang
\address{SUBATECH, UMR 6457, Universit\'e de Nantes, Ecole des Mines de Nantes,
IN2P3/CRNS. 4 rue Alfred Kastler, 44307 Nantes cedex 3, France}
}

\maketitle
\begin{abstract}
We study nonequilibrium effects at the QCD phase transition within the framework of Polyakov loop extended chiral 
fluid dynamics. The quark degrees of freedom act as a locally equilibrated heat bath for the sigma field and a dynamical 
Polyakov loop. Their evolution is described by a Langevin equation with dissipation and noise. At a critical point we observe the formation of long-range 
correlations after equilibration. During a 
hydrodynamical expansion nonequilibrium fluctuations are enhanced at the first order phase transition compared to the critical point.
\end{abstract}
\PACS{25.75.-q, 47.75.+f, 24.60.Ky, 25.75.Nq}
  
\section{Introduction}
Presently the knowledge of the QCD phase diagram is still limited. While lattice QCD calculations tell us 
that at vanishing baryochemical potential there is a crossover \cite{Aoki:2006we}, only model studies allow to 
explore the high baryon density regions. There one expects a first order transition at large $\mu$ ending
at a critical point (CP) \cite{Scavenius:2000qd}. The CP is expected to be detected in heavy-ion collisions 
through event-by-event fluctuations of quantities like transverse momentum or particle multiplicity 
\cite{Stephanov:1998dy,Stephanov:1999zu}. Nevertheless, phenomena like critical slowing down as well as the finite 
system lifetime and size prevent the correlation length from diverging \cite{Berdnikov:1999ph} which thus weakens 
these signals. It is therefore important to include nonequilibrium effects and the dynamics of
the system to estimate the experimental signatures of the critical end point.
A promising ansatz to study the QCD phase transition in such a setting is
provided by the framework of chiral fluid dynamics \cite{Mishustin:1998yc,Paech:2003fe,Nahrgang:2011mg}. 
Here, the basic
idea is to propagate the order parameter of chiral symmetry explicitly by
a Langevin equation, while the heat bath is given by a fluid dynamically
expanding medium made out of quarks. Presently, we extend this model
with the Polyakov loop to consider both the chiral and the deconfinement
transition. Here, the Polyakov loop is treated as an effective field which is
propagated by a phenomenological Langevin equation. We present results
of temperature quenches in a box and study the evolution of fluctuations
during the expansion of a hot plasma droplet.

\section{Chiral fluid dynamics with a Polyakov loop}
\begin{figure}[htbp]
\centering
\begin{minipage}[t]{0.45\textwidth}
\epsfig{file=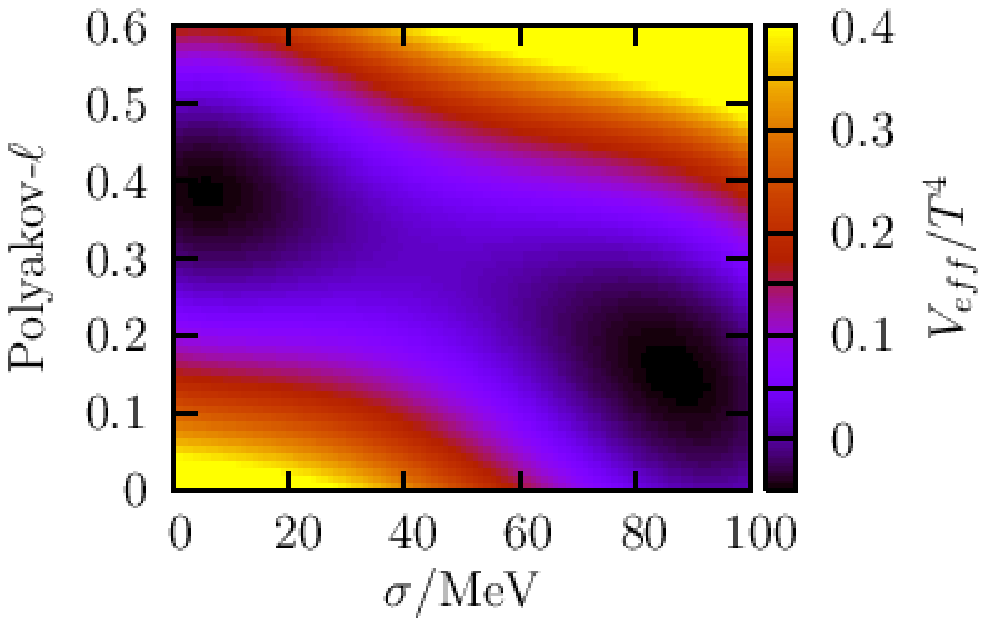, scale=0.5}
\caption{Effective potential at $g=4.7$, first order, $T_c=172.9 MeV$.}
\label{fig:fopot}
\end{minipage}
\hfill
\begin{minipage}[t]{0.45\textwidth}
\epsfig{file=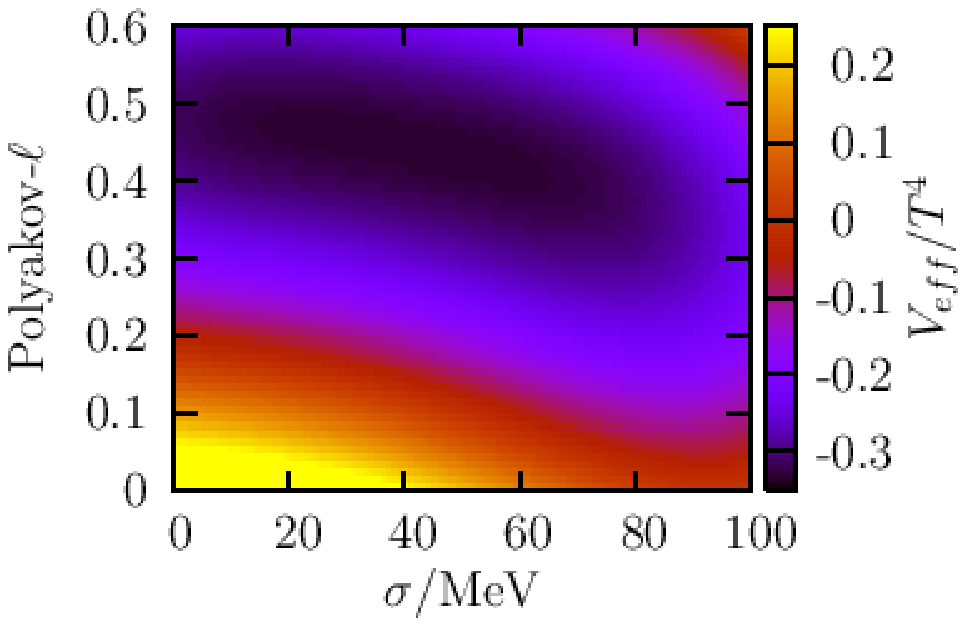, scale=0.5}
\caption{Effective potential at $g=3.52$, CP, $T_c=180.5 MeV$.}
\label{fig:cppot}
\end{minipage}
\end{figure}

For investigation we use the Polyakov loop extended quark meson model \cite{arXiv:0704.3234} with the Lagrangian
\begin{equation}
{\cal L}=\overline{q}\left[i \left(\gamma^\mu \partial_\mu-i g_{\rm QCD}\gamma^0 A_0\right)-g \sigma \right]q + \frac{1}{2}\left(\partial_\mu\sigma\right)^2 
- U\left(\sigma\right) - {\cal U}(\ell, \bar\ell)~,
 \end{equation}
where $q=(u,d)$ is the constituent quark field, $A_0$ the temporal component of the color gauge field and $\sigma$ the mesonic 
field. The pion degrees of freedom are neglected in the present study. The potential for the sigma field 
reads
 \begin{equation}
U\left(\sigma\right)=\frac{\lambda^2}{4}\left(\sigma^2-\nu^2\right)^2-h_q\sigma-U_0~.
\end{equation}
The temperature dependent Polyakov loop potential is chosen in a polynomial form \cite{arXiv:0704.3234, Ratti:2005jh} following
\begin{equation}
\frac{{\cal U}}{T^4}\left(\ell, \bar\ell\right)= -\frac{b_2(T)}{4}\left(\left|\ell\right|^2+\left|\bar\ell\right|^2\right)-\frac{b_3}{6}\left(\ell^3+\bar\ell^3\right) + \frac{b_4}{16}\left(\left|\ell\right|^2+\left|\bar\ell\right|^2\right)^2~.
\end{equation}
Integrating out the quark degrees of freedom, which will constitute the heat bath, we obtain the grand canonical potential. At $\mu_B=0$, $\ell= \bar \ell$ and 
in the mean-field approximation it is \cite{arXiv:0704.3234}
\begin{equation}
 \Omega_{\bar q q}=-4 N_f T\int\frac{\mathrm d^3 p}{(2\pi)^3} \ln\left[1+3\ell\mathrm e^{-\beta E}+3\ell\mathrm e^{-2\beta E}+\mathrm e^{-3\beta E}  \right]~.
\end{equation}
For different coupling strengths $g$ the effective potential $V_{eff}= U+{\cal U}+\Omega_{\bar q q}$
shows different characteristic shapes at the transition temperature: 
for $g=4.7$ we see two degenerate minima, see figure \ref{fig:fopot}, while for $g=3.52$ the potential is very broad and flat at the minimum, figure \ref{fig:cppot}. This 
resembles the situation at a CP. These choices of the quark-meson coupling are of course unphysical but allow a first 
qualitative study of effects at several types of transition
\footnote{Note that in principle one has to choose $g$ such that $g\sigma$ reproduces the constituent
quark mass in vacuum. This would give a value of $g \sim 3.2$.}.

In \cite{Nahrgang:2011mg} the coupled dynamics for the sigma field and the quark fluid were derived self-consistently. The sigma field is propagated by a Langevin equation 

\begin{equation}
 \partial_{\mu}\partial^{\mu}\sigma+\eta_{\sigma}\partial_t \sigma+\frac{\partial V_{\rm eff}}{\partial\sigma}=\xi_{\sigma}\, .
\end{equation}

The explicit form of the temperature dependent damping coefficient $\eta_{\sigma}$ together with the correlator for 
the stochastic noise field have been derived in \cite{Nahrgang:2011mg}.

We allow for a dynamical evolution of the Polyakov loop by adding a kinetic term in the equation of motion \cite{Dumitru:2001}. For the full 
nonequilibrium description we also need to add a damping term $\eta_{\ell}\sim 1/fm$ \cite {Dumitru:2002} and impose the 
dissipation-fluctuation relation:
\begin{equation}
  \frac{2N_c}{g_{\rm QCD}^2}\partial_{\mu}\partial^{\mu}\ell T^2+\eta_{\ell}\partial_t \ell+\frac{\partial V_{\rm eff}}{\partial\ell}=\xi_{\ell}
\end{equation}
\begin{equation}
 \langle\xi_{\ell}(t)\xi_{\ell}(t')\rangle=\frac{1}{V}\delta(t-t')2\eta_{\ell} T~.
\end{equation} 
Note at this point that the Polyakov loop is originally defined only in equilibrium and it is not a priori clear what the correct dynamics are \cite{Ratti:2005jh}.
This approach is, therefore, purely phenomenological.

The quark fluid is propagated via the equations of ideal relativistic fluid dynamic
using the energy-momentum tensors of the liquid, the $\sigma$-field and
the Polyakov loop

\begin{equation}
 \partial_{\mu}\left(T^{\mu\nu}_q+T^{\mu\nu}_{\sigma}+T^{\mu\nu}_{\ell}\right)=0~.
\label{eq:fluiddynamics}
\end{equation}

\section{Numerical results}

\subsection{Temperature quench in a box}
\begin{figure}[htbp]
\centering
\begin{minipage}[t]{0.45\textwidth}
\epsfig{file=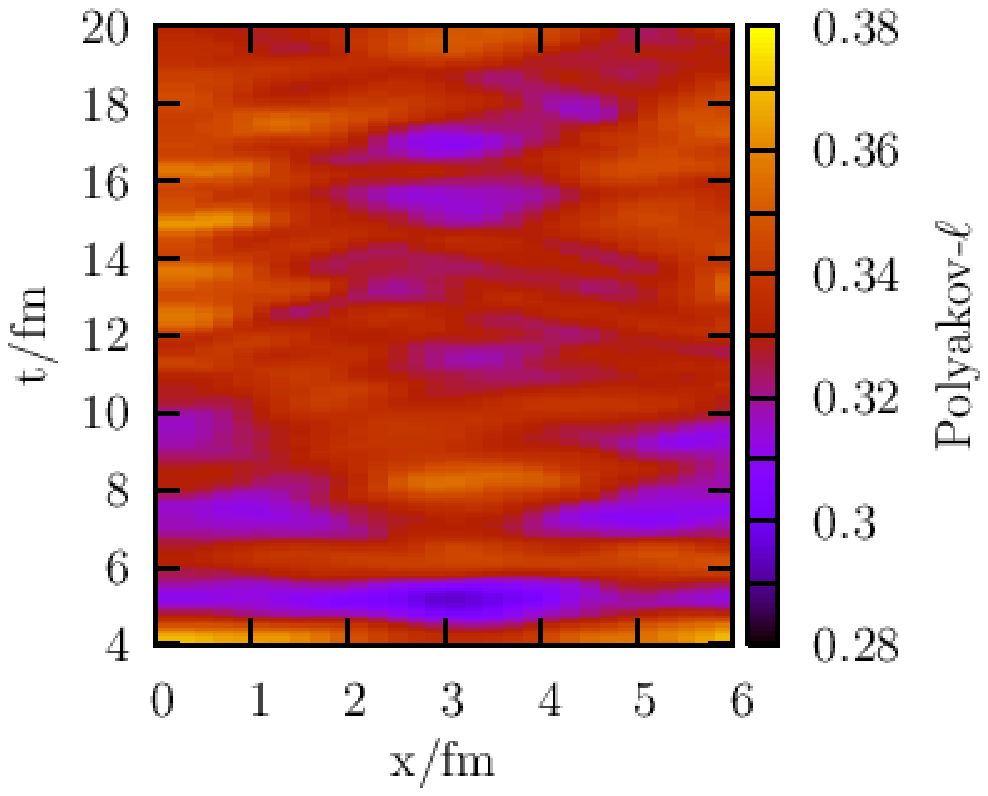, scale=0.5}
\caption{Fluctuations at the first order transition during relaxation process, quench from $T=180$~MeV to $T=140$~MeV.}
\label{fig:foshort}
\end{minipage}
\hfill
\begin{minipage}[t]{0.45\textwidth}
\epsfig{file=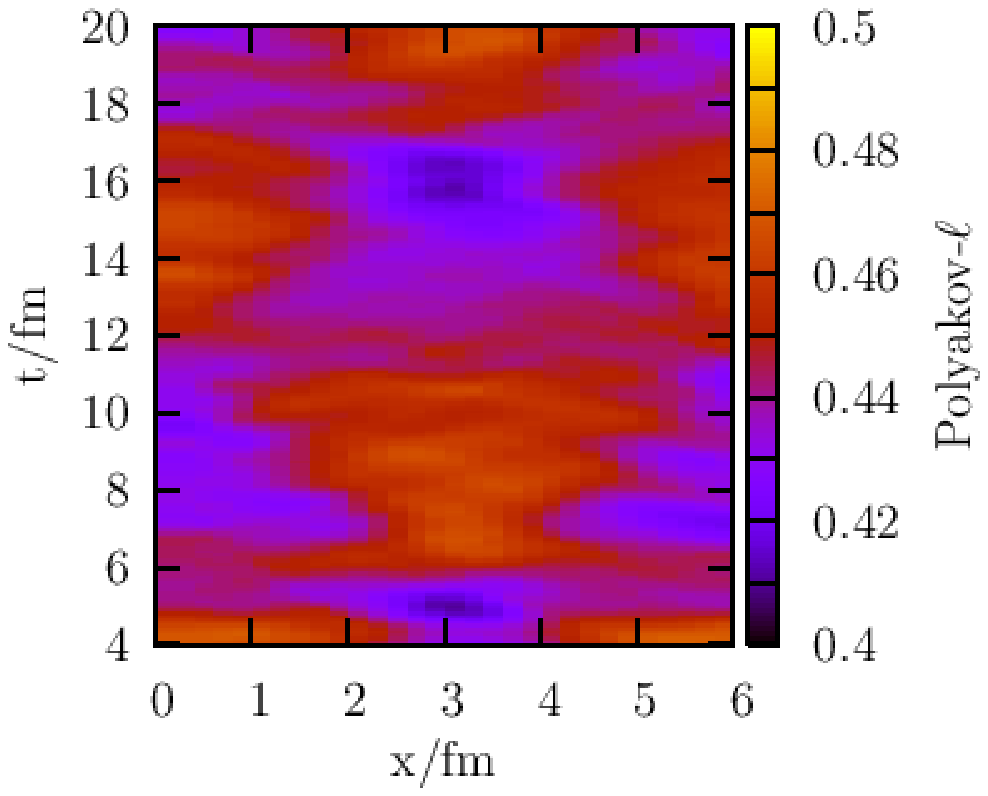, scale=0.5}
\caption{Long-range fluctuations at the CP after relaxation process, quench from $T=186$~MeV to $T=166$~MeV.}
\label{fig:cplong}
\end{minipage}
\end{figure}

We put fields and fluid in a box with periodic boundary conditions. The fields are initialized at a temperature above 
$T_c$. Then the temperature is quenched to a value below the transition point and the quark bath is added. 
The fields lose energy through damping, transferring this amount of energy to the fluid via equation 
(\ref{eq:fluiddynamics}) which leads to a subsequent increase of the 
temperature, see also \cite{Nahrgang:2011ll}. We find quench temperatures such that the system relaxes near the transition temperature. 
In the Polyakov loop, we can observe an interesting phenomenon. Figures \ref{fig:foshort} and \ref{fig:cplong} show 
the value of $\ell$ in $x$-direction $y = z = 0$ evolving in time for both transition scenarios during
the relaxation process. At the CP, one observes the formation of long-range fluctuations over both space 
and time, an effect that does not occur at the first order transition.

\subsection{Fluid dynamic expansion}
\begin{figure}[htb!]
\centering
\epsfig{file=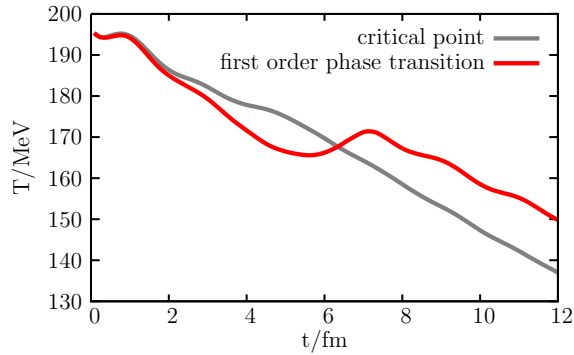, scale = 0.6}
\caption{Temperature of the quark fluid, for the first order scenario reheating after formation of a supercooled phase is observed.}
\label{fig:evolTemp}
\end{figure}

For this simulation we initialize an ellipsoidal region with a temperature of $T=200$~MeV, smoothed by a Woods-Saxon-function 
at the edges. The fields and fluid are set to their equilibrium values throughout the lattice. We let the 
system expand by full $3+1$ dimensional fluid dynamics and measure the average temperature in a fixed central volume as a function of time. The result is shown 
in figure \ref{fig:evolTemp}. While for the CP $T$ decreases monotonically with time, we can observe a reheating at 
the first order transition as a consequence of the forming of a supercooled phase below the transition temperature that finally decays 
to the global minimum and transfers its energy into the fluid.

\begin{figure}[htbp]
\centering
\begin{minipage}[b]{0.45\textwidth}
\epsfig{file=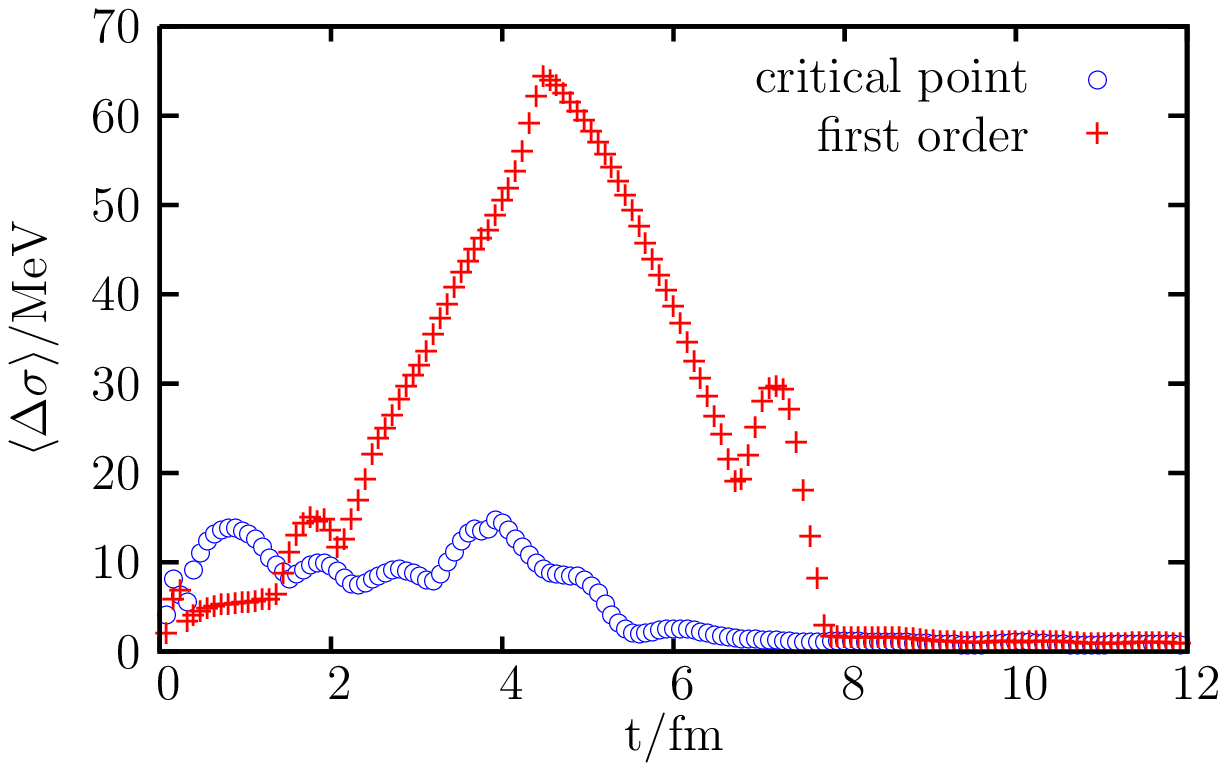, scale=0.4}
\caption{Sigma fluctuations grow large at the first order transition compared to the CP.}
\label{fig:sigmafluct}
\end{minipage}
\hfill
\begin{minipage}[b]{0.45\textwidth}
\epsfig{file=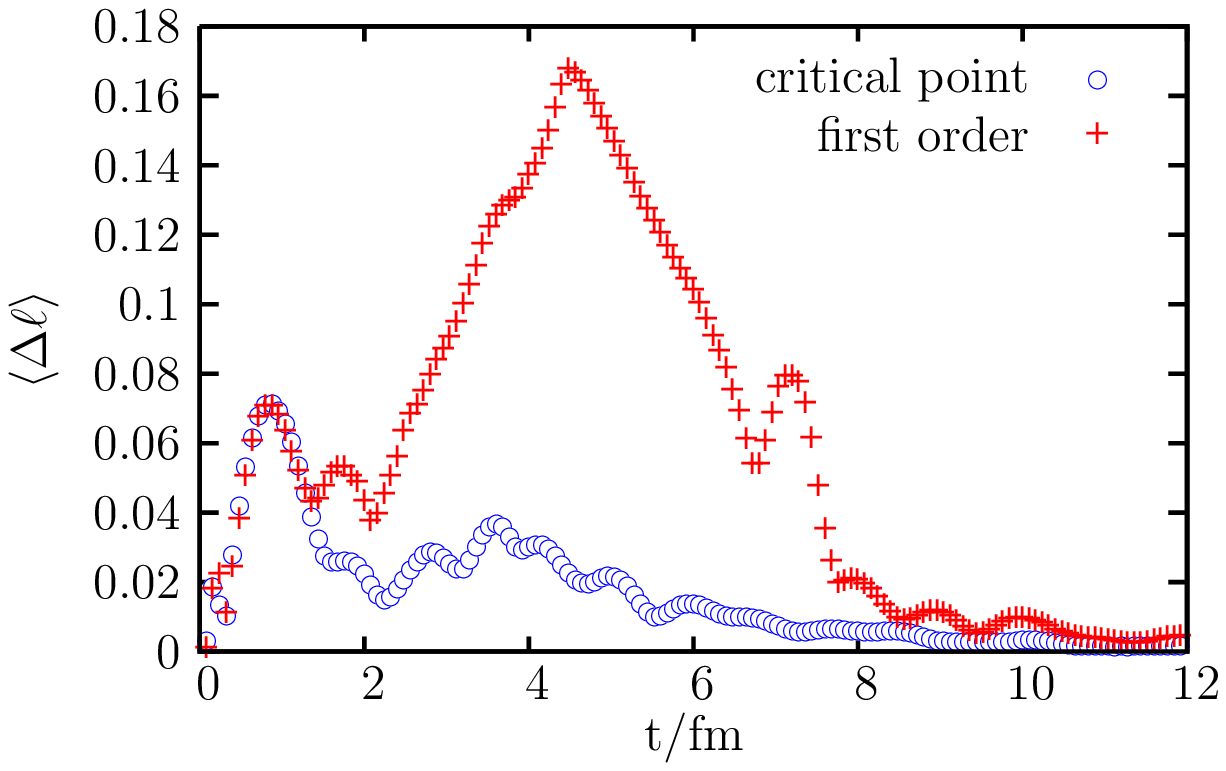, scale=0.4}
\caption{Polyakov loop fluctuations grow large at the first order transition compared to the CP.}
\label{fig:loopfluct}
\end{minipage}
\end{figure}

In figures \ref{fig:sigmafluct} and \ref{fig:loopfluct} we show the evolution of nonequilibrium fluctuations for $\sigma$ and $\ell$.
These are defined as $\langle\Delta\sigma\rangle=\sqrt{\langle\left(\sigma-\sigma_{\rm eq}\right)^2\rangle}$ and $\langle\Delta\ell\rangle=\sqrt{\langle\left(\ell-\ell_{\rm eq}\right)^2\rangle}$, respectively.
For both order parameters we find a large increase of the nonequilibrium fluctuations in a scenario with a the first order transition compared to the scenario with a CP. This is again 
caused by the large deviations from equilibrium that occur during supercooling. We see in both figures a second smaller 
increase, this arises when parts of the system cross the transition temperature a second time after the 
reheating.

\section{Conclusions}

We presented a dynamical model to study the chiral and deconfinement phase transition of QCD. The nonequilibrium evolution
of both order parameters and their interaction with the quark fluid was described by Langevin equations. At the critical 
point we observed the formation of long-range fluctuations. During the hydrodynamical expansion of a finite size system 
our model shows nonequilibrium effects like supercooling and reheating of the quark heat bath at the first order phase 
transition. There we also found an enhancement of nonequilibrium fluctuations in a first order phase transition scenario 
compared to an evolution through the CP.

This work was supported by the Hessian LOEWE initiative Helmholtz International Center for FAIR.


\begin{thebibliography}{50}
\bibitem{Aoki:2006we}
  Y.~Aoki, G.~Endrodi, Z.~Fodor, S.~D.~Katz, K.~K.~Szabo,
  Nature {\bf 443 } (2006)  675-678.

\bibitem{Scavenius:2000qd}
  O.~Scavenius, A.~Mocsy, I.~N.~Mishustin and D.~H.~Rischke,
  Phys.\ Rev.\  C {\bf 64} (2001) 045202.

\bibitem{Stephanov:1998dy}
  M.~A.~Stephanov, K.~Rajagopal and E.~V.~Shuryak,
  Phys.\ Rev.\ Lett.\  {\bf 81} (1998) 4816

\bibitem{Stephanov:1999zu}
  M.~A.~Stephanov, K.~Rajagopal and E.~V.~Shuryak,
  Phys.\ Rev.\  D {\bf 60} (1999) 114028

\bibitem{Berdnikov:1999ph}
  B.~Berdnikov and K.~Rajagopal,
  Phys.\ Rev.\  D {\bf 61}, 105017 (2000)

\bibitem{Mishustin:1998yc}
  I.~N.~Mishustin and O.~Scavenius,
  Phys.\ Rev.\ Lett.\  {\bf 83} (1999) 3134

\bibitem{Paech:2003fe}
  K.~Paech, H.~Stoecker and A.~Dumitru,
  Phys.\ Rev.\  C {\bf 68} (2003) 044907

\bibitem{Nahrgang:2011mg}
  M.~Nahrgang, S.~Leupold, C.~Herold, M.~Bleicher,
  Phys.\ Rev.\  {\bf C84 } (2011)  024912.

\bibitem{arXiv:0704.3234}
  B.~-J.~Schaefer, J.~M.~Pawlowski and J.~Wambach,
  Phys.\ Rev.\ D\ {\bf 76} (2007) 074023

\bibitem{Ratti:2005jh}
  C.~Ratti, M.~A.~Thaler, W.~Weise,
  Phys.\ Rev.\  {\bf D73 } (2006)  014019.

\bibitem{Dumitru:2001}
  A.~Dumitru and R.~D.~Pisarski,
  Phys.\ Lett.\ B\ {\bf 504} (2001) 282

\bibitem{Dumitru:2002}
  A.~Dumitru and R.~D.~Pisarski,
  Nucl.\ Phys.\ A\ {\bf 698} (2002) 444

\bibitem{Nahrgang:2011ll}
  M.~Nahrgang, S.~Leupold, M.~Bleicher,
   [arXiv:1105.1396 [nucl-th]].



\end{thebibliography}
\end{document}